\documentclass[notitlepage,nofootinbib,preprintnumbers,amssymb,superscriptaddress]{revtex4-1}
\usepackage{amsfonts,amssymb,mathtools,graphicx,color,bm}
\definecolor{ultramarine}{rgb}{0.07, 0.04, 0.56}
\definecolor{cadmiumgreen}{rgb}{0.0, 0.42, 0.24}
\definecolor{indigo(dye)}{rgb}{0.0, 0.25, 0.42}
\usepackage[linktocpage=true]{hyperref}
\hypersetup{
colorlinks=true,
citecolor=ultramarine,
linkcolor=cadmiumgreen,
urlcolor=indigo(dye),
}

\usepackage{autobreak}
\newcommand{\D}{{\rm d}}
\newcommand{\fr}[2]{\frac{#1}{#2}}
\newcommand{\pa}{\partial}

\newcommand{\na}{\nabla}
\newcommand{\bra}[1]{\left( #1 \right)}  
\newcommand{\brb}[1]{\left[ #1 \right]}  
\newcommand{\brc}[1]{\left\{ #1 \right\}}  
\newcommand{\be}{\begin{equation}}  
\newcommand{\ee}{\end{equation}}
\newcommand{\bem}{\begin{bmatrix}}
\newcommand{\eem}{\end{bmatrix}}
\newcommand{\Mpl}{M_{\rm Pl}}

\newcommand{\ga}{\gamma}

\newcommand{\la}{\lambda}
\newcommand{\si}{\sigma}

\newcommand{\mn}{{\mu \nu}}
\newcommand{\mE}{\mathcal{E}}

\newcommand{\mO}{\mathcal{O}}
\newcommand{\mR}{\mathcal{R}}

\newcommand{\cU}{c_{{\rm U}}}
\newcommand{\cD}{c_{{\rm D}1}}
\newcommand{\cDD}{c_{{\rm D}2}}

\begin{document}

\preprint{YITP-22-161, IPMU22-0071}

\title{Approximately stealth black hole in higher-order scalar-tensor theories}

\author{Antonio De Felice}
\affiliation{Center for Gravitational Physics and Quantum Information, Yukawa Institute for Theoretical Physics, Kyoto University, Kyoto 606-8502, Japan}

\author{Shinji Mukohyama}
\affiliation{Center for Gravitational Physics and Quantum Information, Yukawa Institute for Theoretical Physics, Kyoto University, Kyoto 606-8502, Japan}
\affiliation{Kavli Institute for the Physics and Mathematics of the Universe (WPI), The University of Tokyo Institutes for Advanced Study, The University of Tokyo, Kashiwa, Chiba 277-8583, Japan}

\author{Kazufumi Takahashi}
\affiliation{Center for Gravitational Physics and Quantum Information, Yukawa Institute for Theoretical Physics, Kyoto University, Kyoto 606-8502, Japan}

\begin{abstract}
We investigate a generic quadratic higher-order scalar-tensor theory with a scordatura term, which is expected to provide a consistent perturbative description of stealth solutions with a timelike scalar field profile. In the DHOST subclass, exactly stealth solutions are known to yield perturbations infinitely strongly coupled and thus cannot be trusted. Beyond DHOST theories with the scordatura term, such as in ghost condensation and U-DHOST, we show that stealth configurations cannot be realized as exact solutions but those theories instead admit approximately stealth solutions where the deviation from the exactly stealth configuration is controlled by the mass scale~$M$ of derivative expansion. The approximately stealth solution is time-dependent, which can be interpreted as the black hole mass growth due to the accretion of the scalar field. From observed astrophysical black holes, we put an upper bound on $M$ as $\hat{c}_{\rm D1}^{1/2} M\lesssim 2\times 10^{11}$~GeV, where $\hat{c}_{\rm D1}$ is a dimensionless parameter of order unity that characterizes the scordatura term. As far as $M$ is sufficiently below the upper bound, the accretion is slow and the approximately stealth solutions can be considered as stealth at astrophysical scales for all practical purposes while perturbations are weakly coupled all the way up to the cutoff~$M$ and the apparent ghost is as heavy as or heavier than $M$. 
\end{abstract}

\maketitle

\section{Introduction}\label{sec:intro}

We are living in very exciting times in cosmology. New and precise experiments put the $\Lambda$CDM concordance model to the test. These experiments span a large range of cosmic history and redshifts. At the moment, we see tensions in the data when we want to explain the Planck data (i.e., high-redshift data) and late-time (low-redshift) data, 
when we use the $\Lambda$CDM model to fit all these data sets at the same time. In particular, what is now known as the $H_0$ tension, unless some unknown systematics are found, puts serious constraints on the concordance model of cosmology. As if this is not enough, there could be other tensions in the data regarding the structure formation, again, once we compare the predictions of Planck-$\Lambda$CDM with the most recent observations of large-scale structure. These tensions might go away in a few years, after we recognize what is wrong with the experiments or the interpretations/analyses we do on the data.

However, if the tensions are going to be confirmed and made worse by new data and experiments, then we have to abandon the concordance model for something else. A crucial point then would be what direction we should follow to modify the concordance model. At the moment we have seen a plethora of possibilities and some of them could actually end up describing our Universe better than $\Lambda$CDM. However, it should be said that these possibilities usually come as a bottom-up approach. Namely, we do not have a theory, say the theory of everything, which is able to tell us which directions we should be going now, in case $\Lambda$CDM has to be abandoned.
In particular, this seems related to the intrinsically unknown dominant components in our Universe: dark energy and dark matter. They, by themselves, dominate today's evolution of our Universe in which known forms of matter, baryons and radiations, are responsible for only approximately 5\% of the total energy density. The concordance model is based on general relativity (GR) and it gives a description of this dark Universe in terms of a cosmological constant to model dark energy and a cold, pressureless dust component as dark matter. 
The concordance model probably gives the simplest explanation for dark energy/matter.
Moreover, GR passes all solar system constraints and was confirmed in the prediction of the existence of gravitational waves.
Nevertheless, if future data exclude $\Lambda$CDM, we will have to change something in the modeling of the dark Universe, which is one of the main motivations to study modified gravity~\cite{Koyama:2015vza,Ferreira:2019xrr,Arai:2022ilw}.

A simple extension of the concordance model, typically introduced to describe dark energy, is the introduction of a scalar field~$\phi$ at the level of the Lagrangian. Then, once we assume the existence of such a scalar field to be affecting the late-time behavior of the Universe, the question is which Lagrangian one should introduce to describe its dynamics.
In particular, in these last few years of research in modified gravity, there has been an exploration in theory space for scalar-tensor theories which would generalize 
traditional models, such as quintessence, while keeping the theory consistent. One property we tend to believe such theories should possess is the absence of the 
Ostrogradsky ghost appearing when the equations of motion (EOMs) are of third or higher order~\cite{Woodard:2015zca}.
Such a degree of freedom would make the vacuum unstable, and one would need to find a way to get a viable cosmology and even to make predictions in such a theory.
Therefore, people started studying theories without this unwanted degree of freedom. 
The most general class of scalar-tensor theories with second-order Euler-Lagrange equations is called the Horndeski theory~\cite{Horndeski:1974wa,Deffayet:2011gz,Kobayashi:2011nu}, and hence it serves as a general framework for scalar-tensor theories without the Ostrogradsky ghost.
Nevertheless, the Horndeski theory is not the most general class of ghost-free scalar-tensor theories:
Theories could be of higher order and still be free from the Ostrogradsky ghost, provided that their Lagrangian satisfies some degeneracy conditions~\cite{Motohashi:2014opa,Langlois:2015cwa,Motohashi:2016ftl,Klein:2016aiq,Motohashi:2017eya,Motohashi:2018pxg}. Such theories, named as degenerate higher-order scalar-tensor (DHOST) theories~\cite{Langlois:2015cwa,Crisostomi:2016czh,BenAchour:2016fzp,Takahashi:2017pje,Langlois:2018jdg}, 
would have a non-trivial phenomenology that could be tested against the data and which would differ from the concordance model and even from the Horndeski theory 
(see Refs.~\cite{Langlois:2018dxi,Kobayashi:2019hrl} for reviews).
A systematic way to construct DHOST theories is the invertible disformal transformation~\cite{Bekenstein:1992pj} or its higher-derivative generalization~\cite{Takahashi:2021ttd} since an invertible transformation preserves the number of physical degrees of freedom~\cite{Domenech:2015tca,Takahashi:2016dnv}.
The most general class of DHOST theories to this date is obtained by performing the generalized disformal transformation on the Horndeski theory and is called the generalized disformal Horndeski class~\cite{Takahashi:2022mew}.
One could further enlarge the theory space of ghost-free scalar-tensor theories by requiring the degeneracy conditions for the higher-derivative terms holding only in the unitary gauge where the scalar field is spatially uniform, and the resultant framework is called U-DHOST~\cite{DeFelice:2018ewo,DeFelice:2021hps}.
For a generic choice of coordinates away from the unitary gauge, there is an apparent Ostrogradsky mode, but it does not actually propagate because it satisfies a three-dimensional elliptic differential equation on a spacelike hypersurface.
Such a mode is referred to as a shadowy mode, which itself is harmless~\cite{DeFelice:2018ewo}.

One problem with DHOST theories is that they do not provide a consistent perturbative description of so-called stealth solutions, i.e., those with the same metric configuration as in GR accompanied by a nontrivial scalar profile.
The first stealth black hole solution was found in \cite{Mukohyama:2005rw} in the context of k-essence, as a precursor of an approximate stealth solution in ghost condensation (with a scordatura term that we shall describe below) found also in the same paper, by using the Lema\^{i}tre reference frame in the Schwarzschild spacetime.
The construction was then extended to Schwarzschild--(anti-)de Sitter in the context of Horndeski theory~\cite{Babichev:2013cya,Kobayashi:2014eva} and DHOST theory~\cite{BenAchour:2018dap,Motohashi:2019sen}.
Moreover, the existence conditions for general stealth solutions in generic higher-order scalar-tensor theories were clarified in \cite{Motohashi:2018wdq,Takahashi:2020hso}.
Although DHOST theories can accommodate stealth solutions, it turned out that their perturbations exhibit strong coupling~\cite{Babichev:2018uiw,deRham:2019gha,Motohashi:2019ymr,Khoury:2020aya,Takahashi:2021bml}.
This problem can be circumvented by taking into account a small detuning of the above mentioned degeneracy conditions, dubbed as ``scordatura'' in \cite{Motohashi:2019ymr}, as already implemented in ghost condensation~\cite{ArkaniHamed:2003uy,Arkani-Hamed:2003juy}.
Since the degeneracy conditions are not protected in general by any symmetry, we should expect quantum corrections to detune them in any case.
Of course, this detuning revives the Ostrogradsky ghost, but from the viewpoint of effective field theory (EFT), the ghost is harmless so long as its mass is above the EFT cutoff.
Cosmology in the modified DHOST theories with scordatura terms was then explored in \cite{Gorji:2020bfl,Gorji:2021isn}. 
Later on, it was shown that U-DHOST theories have the scordatura mechanism 
implemented by default~\cite{DeFelice:2022xvq}, making these theories a safe framework for studying stealth solutions (or approximately stealth solutions, as we will see in this paper). In the same paper~\cite{DeFelice:2022xvq}, it was also shown that if we do consider generic higher-order scalar-tensor theories beyond U-DHOST, on assuming reasonable scaling properties for functions in the Lagrangian in terms of the cutoff scale~$M\,(\ll \Mpl$), then the mass of the Ostrogradsky mode would be of order~$M$ or larger. 
This shows that those theories would represent healthy EFTs able to describe physical phenomena below the cutoff. Above the cutoff, one would need to assume the existence of a good ultraviolet completion that would cure/remove the apparent Ostrogradsky ghost.

In the present paper, we study stealth (or approximately stealth) solutions in generic higher-order scalar-tensor theories, which include ghost condensation, U-DHOST, and DHOST theories.
In particular, we show that these theories (except the DHOST case) do not admit the stealth Schwarzschild(-de Sitter) configuration as an exact solution. Nevertheless, on assuming that $M\gg\Mpl^2/M_{\rm BH}$ (a condition which holds for typical astrophysical black holes with mass~$M_{\rm BH}$ and high-energy, below Planck, cutoff scales), we find 
approximately stealth solutions. The correction to the exact stealth configuration in general depends both on time and space. We will show that, for large values of (the advanced null) time, while keeping the distance from the horizon fixed, 
the correction can be understood as the accretion of the scalar field on the black hole, a phenomenon already found in \cite{Mukohyama:2005rw,Cheng:2006us} in the context of (ungauged and gauged) ghost condensation.
Indeed, analogously to the case of \cite{Mukohyama:2005rw,Cheng:2006us}, we find a constraint coming from the fact that the mass growth due to 
the deviation from DHOST theories should not exceed the difference between the measured mass of a black hole and its initial mass. In particular, we find an upper bound for the cutoff of the theory given by $\hat{c}_{\rm D1}^{1/2}M\lesssim 2\times 10^{11}$~GeV, where $\hat{c}_{\rm D1}$ is a dimensionless parameter of order unity characterizing the deviation from DHOST theories.

The rest of this paper is organized as follows.
In \S\ref{sec:stealth}, we define the theory and show that, unless we have DHOST theories, the stealth Schwarzschild(-de Sitter) configuration is not compatible with the EOMs. In \S\ref{sec:approx_stealth}, we build up approximately stealth solutions for generic higher-order scalar-tensor theories. We find that these approximate solutions lead to a growth in time for the mass of the black hole, and we discuss the phenomenon and its bound on the cutoff scale~$M$ for the theory in \S\ref{sec:BH_accretion}. Finally, we draw our conclusions in \S\ref{sec:conc}.

\section{Stealth solutions in higher-order scalar-tensor theories}\label{sec:stealth}

\subsection{The model}

We consider the shift-symmetric subclass of quadratic higher-order scalar-tensor theories described by the following action:
	\be
	S=\int \D^4x\sqrt{-g}\brb{F_0(X)+F_1(X)\Box \phi+F_2(X)R+\sum_{I=1}^{5}A_I(X)L_I^{(2)}}, \label{HOST}
	\ee
where the coefficients~$F_0$, $F_1$, $F_2$, and $A_I$'s are functions of $X\coloneqq \phi_\mu\phi^\mu$ and 
	\be
	L_1^{(2)}\coloneqq \phi^{\mn}\phi_{\mn}, \quad
	L_2^{(2)}\coloneqq (\Box\phi)^2, \quad
	L_3^{(2)}\coloneqq \phi^\mu\phi_{\mn}\phi^\nu\Box\phi, \quad
	L_4^{(2)}\coloneqq \phi^\mu\phi_{\mn}\phi^{\nu\la}\phi_\la, \quad
	L_5^{(2)}\coloneqq (\phi^\mu\phi_{\mn}\phi^\nu)^2,
	\ee
with $\phi_\mu\coloneqq \na_\mu\phi$ and $\phi_{\mn}\coloneqq \na_\mu\na_\nu\phi$.
For a generic choice of the coefficient functions, the action~\eqref{HOST} yields an Ostrogradsky ghost.
In order to remove the ghost, the higher-derivative terms in the Lagrangian must be chosen to satisfy some degeneracy conditions.
Depending on whether we require the degeneracy in an arbitrary coordinate system or only under the unitary gauge where $\phi=\phi(t)$, we obtain DHOST or U-DHOST theories, respectively.
The degeneracy conditions for DHOST theories are summarized as
    \begin{equation}
    {c_{\rm D1}}={c_{\rm D2}}={c_{\rm U}}=0. \label{DC}
    \end{equation}
Here, we have defined
    \begin{equation}
    \begin{split}
    {c_{\rm D1}}&\coloneqq X(A_1+A_2), \\
    {c_{\rm D2}}&\coloneqq X^2A_4+4X\Upsilon F_{2X}-2{\mathcal{G}}_T+2(1-\Upsilon)\tilde{\mathcal{G}}_T, \\
    {c_{\rm U}}&\coloneqq -X^3A_5-X^2(A_3+A_4)-{c_{\rm D1}}+3\Upsilon^2(2{\mathcal{G}}_T+3{c_{\rm D1}}),
    \end{split}
    \end{equation}
with a subscript $X$ denoting the $X$-derivative and
    \begin{equation}
    {\mathcal{G}}_T\coloneqq F_2-XA_1, \qquad
    \tilde{\mathcal{G}}_T\coloneqq (1-\Upsilon)F_2-2XF_{2X}, \qquad
    \Upsilon\coloneqq -\frac{X(4F_{2X}+2A_2+XA_3)}{2(2{\mathcal{G}}_T+3{c_{\rm D1}})}.
    \label{GTUp}
    \end{equation}
On the other hand, the degeneracy condition for U-DHOST theories is given by $\cU=0$~\cite{Langlois:2015cwa,DeFelice:2018ewo,DeFelice:2021hps,DeFelice:2022xvq}.
A simple model of U-DHOST is given by
    \be
    S=\int \D^4x\sqrt{-g}\brb{\fr{\Mpl^2}{2}R+F_0(X)+A_2(X)\bra{\Box\phi-\fr{1}{X}\phi^\mu\phi_\mn\phi^\nu}^2},
    \ee
with $A_2\ne 0$, which amounts to
    \be
    F_2=\fr{\Mpl^2}{2}, \qquad
    F_1=A_1=A_4=0, \qquad
    A_2=-\fr{X}{2}A_3=X^2A_5. \label{simple_example}
    \ee
One can easily check that this choice of functions satisfies $\cU=0$ and $\cD=XA_2\ne 0$ (and $\cDD=0$).

In what follows, unless otherwise stated, we focus on generic higher-order scalar-tensor theories described by the action~\eqref{HOST} that can in general lie outside the (U-)DHOST class.

\subsection{Existence conditions for stealth solutions}

In this subsection, we derive a set of conditions for the theory~\eqref{HOST} to admit the Schwarzschild(-de Sitter) black hole with a linearly time-dependent scalar hair and a constant $X$ as an exact solution, i.e.,
    \be
    \begin{split}
    &g_\mn^{(0)} \D x^\mu \D x^\nu
    =-A(r)\D t^2+\fr{\D r^2}{A(r)}+r^2\gamma_{ab}\D x^a \D x^b, \qquad
    A(r)=1-\fr{r_s}{r}-\fr{\Lambda}{3}r^2, \\
    &\phi^{(0)}=qt+\psi(r), \qquad
    X=X_0=-q^2,
    \end{split} \label{Sch-dS}
    \ee
where $\gamma_{ab}$ is the metric on a two-dimensional sphere.
Note that we have assumed that the scalar field has a timelike profile, hence $X<0$.
The radial profile of the scalar field is written explicitly as
    \be
    \psi(r)=q\int \fr{\sqrt{1-A}}{A} \D r.
    \ee
Let us introduce two other different coordinate systems which we will use in the following discussion.
In the Painlev{\'e}-Gullstrand (PG) coordinates~$(\tau,r,\theta,\varphi)$, the above background configuration can be written as
    \be
    g_\mn^{(0)} \D x^\mu \D x^\nu
    =-A\D\tau^2+2\sqrt{1-A}\,\D\tau\D r+\D r^2+r^2\gamma_{ab}\D x^a \D x^b, \qquad
    \phi^{(0)}=q\tau,
    \label{Sch-dS_PG}
    \ee
where the new time coordinate~$\tau$ is defined by
    \be
    \D\tau=\D t+\fr{\sqrt{1-A}}{A} \D r. \label{tau->tr}
    \ee
In particular, in the case of $\Lambda=0$, $t$ and $\tau$ are related to each other by
    \be
    \tau=t+2\sqrt{r_sr}+r_s\ln\fr{\sqrt{r}-\sqrt{r_s}}{\sqrt{r}+\sqrt{r_s}}.
    \label{t-to-tau}
    \ee
Also, in the Lema{\^i}tre coordinates~$(\tau,\rho,\theta,\varphi)$, we have
    \be
    g^{(0)}_{\mn}\D x^\mu \D x^\nu=
    -\D\tau^2+(1-A)\D\rho^2+r^2\ga_{ab}\D x^a\D x^b, \qquad
    \phi^{(0)}=q\tau, \label{Sch-dS_Lemaitre}
    \ee
where the new radial coordinate~$\rho$
is defined by
    \be
    \D\rho=\D\tau+\fr{\D r}{\sqrt{1-A}}. \label{rho->tr}
    \ee
This equation implies that the original radial coordinate~$r$ is a function of $\rho-\tau$.

Let us consider vacuum solutions in theories described by the action~\eqref{HOST}.
Thanks to general covariance of the action, the EOM for the scalar field is automatically satisfied if the EOM for the metric is satisfied.
Under the condition~$X=X_0={\rm const}$, the EOM for the metric takes the form
    \begin{align}
    0=\mE_\mn\coloneqq& 2F_2 R_{\mu\nu}
    -\brc{F_0+R F_2 +A_1\phi_{\alpha\beta}^2 -A_2\brb{(\Box\phi)^2-2\phi_{\alpha\beta}^2- 2\phi^\alpha \phi^\beta R_{\alpha\beta} } 
    }g_\mn \nonumber \\
    &+\bigl\{2(F_{0X}+R F_{2X})
    +A_3 \phi^\alpha\phi^\beta R_{\alpha\beta} + 2F_{1X}\Box\phi -(A_3+2A_{1X})\brb{(\Box\phi)^2-\phi_{\alpha\beta}^2}
    +2(A_{1X}+A_{2X})(\Box\phi)^2\bigr\}\phi_\mu\phi_\nu \nonumber \\
    &+2A_1\brb{ (\Box\phi)\phi_\mn - \phi_\mu^\la\phi_{\la\nu} - R_{\mu\la\nu\si}\phi^\la\phi^\si} 
    +4A_2R_{\la(\mu}\phi_{\nu)}\phi^\la
    -4(A_1+A_2)\phi_{(\mu}\Box \phi_{\nu)}. \label{EOM1}
    \end{align}
Note that there is no contribution from $A_4$ and $A_5$ thanks to the background profile~$X={\rm const}$.
Then, since we are interested in the stealth configuration~\eqref{Sch-dS}, we can use the vacuum Einstein equation~$G_\mn=-\Lambda g_\mn$ or $R_\mn=\Lambda g_\mn$ to rewrite the above equation as
    \begin{align}
    \mE_\mn=& -\brc{F_0+2\Lambda F_2 +A_1\phi_{\alpha\beta}^2 -A_2\brb{(\Box\phi)^2-2\phi_{\alpha\beta}^2- 2\Lambda X_0 } 
    }g_\mn \nonumber \\
    &+\bigl\{2(F_{0X}+4\Lambda F_{2X})
    +4\Lambda A_2+\Lambda X_0 A_3 + 2F_{1X}\Box\phi -(A_3+2A_{1X})\brb{(\Box\phi)^2-\phi_{\alpha\beta}^2}
    +2(A_{1X}+A_{2X})(\Box\phi)^2\bigr\}\phi_\mu\phi_\nu \nonumber \\
    &+2A_1\brb{ (\Box\phi)\phi_\mn - \phi_\mu^\la\phi_{\la\nu} - R_{\mu\la\nu\si}\phi^\la\phi^\si}
    -4(A_1+A_2)\phi_{(\mu}\Box \phi_{\nu)}. \label{EOM2}
    \end{align}
We note that the following identities hold for the configuration~\eqref{Sch-dS}~\cite{Takahashi:2020hso}:
    \be
    \begin{split}
    (\Box\phi)\phi_\mn- \phi_\mu^\la\phi_{\la\nu}- R_{\mu\la\nu\si}\phi^\la\phi^\si
    &=\Lambda(q^2 g_\mn +\phi_\mu\phi_\nu), \\
    (\Box\phi)^2- \phi_{\alpha\beta}^2
    &=2\Lambda q^2.
    \end{split}
    \ee
By use of these identities, \eqref{EOM2} can be rewritten as
    \begin{align}
    \mE_\mn=& -\brb{F_0+2\Lambda F_2-2\Lambda q^2(A_1+2A_2)+(A_1+A_2)\phi_{\alpha\beta}^2
    }g_\mn \nonumber \\
    &+\bigl\{2(F_{0X}+4\Lambda F_{2X})
    +\Lambda (2A_1-4q^2A_{1X}+4A_2-3q^2A_3)+2F_{1X}\Box\phi
    +2(A_{1X}+A_{2X})(\Box\phi)^2\bigr\}\phi_\mu\phi_\nu \nonumber \\
    &-4(A_1+A_2)\phi_{(\mu}\Box \phi_{\nu)}, \label{EOM3}
    \end{align}
which consists of three tensorial quantities, i.e., $g_\mn$, $\phi_\mu\phi_\nu$, and $\phi_{(\mu}\Box \phi_{\nu)}$.

If the functions~$A_1(X)$ and $A_2(X)$ satisfy $A_1+A_2=0$, then \eqref{EOM3} reduces to
    \be
    \mE_\mn=-\brb{F_0+2\Lambda \bra{F_2+q^2A_1}}g_\mn
    +\brb{2(F_{0X}+4\Lambda F_{2X})
    -\Lambda (2A_1+4q^2A_{1X}+3q^2A_3)+2F_{1X}\Box\phi}\phi_\mu\phi_\nu, \label{EOM4}
    \ee
Hence, the configuration~\eqref{Sch-dS} solves the EOM for the theory~\eqref{HOST} if the following conditions are satisfied at $X=-q^2$:
    \be
    F_0+2\Lambda (F_2+q^2A_1)=0, \qquad
    2(F_{0X}+4\Lambda F_{2X})-\Lambda (2A_1+4q^2A_{1X}+3q^2A_3)=0, \qquad
    F_{1X}=0,
    \ee
or equivalently,
    \be
    \Lambda=-\fr{F_0}{2(F_2+q^2A_1)}, \qquad
    F_0(2A_1+4q^2A_{1X}+3q^2A_3-8F_{2X})+4F_{0X}(F_2+q^2A_1)=0, \qquad
    F_{1X}=0.
    \label{exist}
    \ee
Here, the first two conditions can be used to fix the value of $\Lambda$ and $q^2$.

Now, we consider the case~$A_1+A_2\ne 0$.
Let us write the background configuration in terms of the Lema{\^i}tre coordinates [see Eq.~\eqref{Sch-dS_Lemaitre}].
In this particular coordinate system, both $g_\mn$ and $\phi_\mu\phi_\nu$ are diagonal, while $\phi_{(\mu}\Box \phi_{\nu)}$ has an off-diagonal ($\tau\rho$) component,
    \be
    \phi_{(\tau}\Box \phi_{\rho)}
    =-\fr{q^2}{8r^2(1-A)}\bra{2-2A+r\fr{\D A}{\D r}}^2
    \propto q^2r_s^2.
    \ee
Therefore, the EOM cannot be satisfied unless $q=0$ or $r_s=0$, which is consistent with the result of \cite{Mukohyama:2022skk}.
This shows that theories with $c_{\rm D1}\propto A_1+A_2\ne 0$, which accommodate the scordatura mechanism~\cite{Motohashi:2019ymr,DeFelice:2022xvq}, do not admit the stealth configuration as an exact solution.
In the following subsection, we study an approximate solution for theories with $c_{\rm D1}\ne 0$, where the deviation from the exact stealth configuration is controlled by the parameter $c_{\rm D1}$.

\section{Approximately stealth solution}\label{sec:approx_stealth}

Having clarified that higher-order scalar-tensor theories with $\cD\propto A_1+A_2\ne 0$ do not admit the stealth Schwarzschild(-de Sitter) configuration as an exact solution, let us try to find an approximately stealth solution in theories with $\cD\ne 0$. In what follows, by assuming that $\cD$ is sufficiently small in an appropriate unit that we shall identify below, we consider the term proportional to $\cD$ as a source of small deviation from the stealth Schwarzschild configuration with $\Lambda=0$ and treat such deviation perturbatively in a controlled way. In this case, the background (i.e., zeroth-order in $\cD$) value of $F_0$ and $F_1$ satisfy
    \be
    F_0(-q^2)=0, \qquad
    F_{0X}(-q^2)=0, \qquad
    F_{1X}(-q^2)=0.
    \ee
We will first focus on the simple U-DHOST model~\eqref{simple_example}, and then proceed to the general higher-order scalar-tensor theories described by the action~\eqref{HOST}.
For the model~\eqref{simple_example} with $\cD=XA_2$, assuming the scaling property~$\cD=\hat{c}_{\rm D1}M^2$ and $q=\hat{q}M^2$ where $M\,(\ll\Mpl)$ is the mass scale of derivative expansion and $\hat{c}_{\rm D1}$ and $\hat{q}$ are dimensionless quantities of order unity, one can estimate the effect of the higher-derivative interaction in the action as\footnote{As shown in \cite{DeFelice:2022xvq}, the condition~$\hat{c}_{\rm D1}>0$ is necessary for the scordatura mechanism to be successful.} 
    \be
    0<\fr{\hat{c}_{\rm D1}}{M^6}\bra{\Box\phi-\fr{1}{X}\phi^\mu\phi_\mn\phi^\nu}^2
    =\fr{\hat{c}_{\rm D1}}{M^6}\fr{9q^2r_s}{4r^3}
    <\fr{9\hat{q}^2\hat{c}_{\rm D1}}{4}\fr{1}{M^2r_s^2},
    \ee
where we have substituted the stealth configuration~\eqref{Sch-dS_PG} with $\Lambda=0$.
Note that the factor
    \be
    \epsilon\coloneqq \fr{1}{M^2r_s^2}
    =\fr{16\pi^2\Mpl^4}{M^2M_{\rm BH}^2}
    \ee
is small for typical astrophysical black holes with mass $M_{\rm BH}$.
This suggests that, for the simple model~\eqref{simple_example}, there is an approximately stealth solution whose deviation from the exact stealth configuration is characterized by the small factor~$\epsilon$.
Interestingly, as we shall see below, the solution for the general case~\eqref{HOST} is essentially described by the solution for the simple model of U-DHOST in the sense that any correction to it is suppressed by $M/\Mpl$ and/or $\epsilon$.

Let us consider the stealth configuration~\eqref{Sch-dS_PG} plus perturbations of order~$\epsilon$ in the PG coordinate system, where one can find an analytic solution as we see below.
Spherically symmetric perturbations about the stealth configuration can be written as
	\be
	\begin{split}
	&\delta g_{\tau \tau}=A(r)H_0(\tau,r), \qquad
	\delta g_{\tau r}=\sqrt{1-A}\,H_1(\tau,r), \qquad
	\delta g_{rr}=H_2(\tau,r), \\
	&\delta g_{\tau a}=\delta g_{ra}=0, \qquad
	\delta g_{ab}=r^2K(\tau,r)\ga_{ab}, \qquad
	\delta\phi=\delta\phi(\tau,r),
	\end{split}
	\ee
with $\delta g_\mn\coloneqq g_\mn-g^{(0)}_\mn$ and $\delta \phi\coloneqq \phi-\phi^{(0)}$.
In order for the perturbative description to be valid, the variables~$H_0$, $H_1$, and $H_2$ should be negligibly small compared to unity.
The gauge degrees of freedom can be used to fix $K=0=\delta\phi$, and hence we keep only the three variables~$H_0$, $H_1$, and $H_2$.
Then, the $tt$-, $tr$-, and $rr$-components of the metric EOM provide three independent EOMs for perturbations.
Schematically, the EOMs have the form
    \be
    \begin{split}
    \mE_{tt}[H_0,H_1,H_2]&=\cD(\cdots)-q^2c_{{\rm D}1,X}(\cdots), \\
    \mE_{tr}[H_0,H_1,H_2]&=\cD(\cdots)-q^2c_{{\rm D}1,X}(\cdots), \\
    \mE_{rr}[H_0,H_1,H_2]&=\cD(\cdots)-q^2c_{{\rm D}1,X}(\cdots),
    \end{split} \label{EOM_pert}
    \ee
where the terms with $\cD$ in the right-hand sides are of $\mO(\epsilon)$ and the ellipses denote some functionals of $A\,(=-g_{tt})$ associated with the background Schwarzschild spacetime.
Here, $\mE_{tt}$, $\mE_{tr}$, and $\mE_{rr}$ are homogeneous linear polynomials in $H_0$, $H_1$, $H_2$, and their derivatives up to second order.
The solution to this inhomogeneous system of differential equations is written as a sum of a particular solution sourced by $\cD$ and any solution to the associated homogeneous system of equations.

On finding the solution to \eqref{EOM_pert}, it is useful to define the perturbation variable corresponding to the perturbation of the Misner-Sharp mass.
On the spherically symmetric spacetime, the Misner-Sharp mass is given by
    \be
    1-\fr{2M_{\rm MS}(\tau,r)}{r}
    =g^\mn\pa_\mu r\pa_\nu r
    \eqqcolon 1-\fr{r_s(1+\delta)}{r},
    \ee
where
    \be
    \delta=\bra{1-\fr{r_s}{r}}\brb{H_0+2H_1+\bra{\fr{r}{r_s}-1}H_2}. \label{delta}
    \ee
In what follows, we regard $\delta$, $H_1$, and $H_2$ as independent variables.
In order for the perturbative description to be valid, we require that these perturbation variables do not diverge in the limit~$r\to \infty$.
Regarding $H_2$, it should vanish at least as fast as $r^{-1}$, as otherwise $\delta$ would diverge at the spatial infinity [see Eq.~\eqref{delta}]. We shall see below that these conditions uniquely determine the solution of the inhomogeneous system of differential equations~\eqref{EOM_pert}.

As mentioned earlier, we first focus on the simple U-DHOST model~\eqref{simple_example}, for which $\cD=XA_2$.
By taking an appropriate linear combination of the EOMs~\eqref{EOM_pert}, we find the following differential equation for $\delta$:
    \be
    \bra{\pa_r-\sqrt{\fr{r}{r_s}}\pa_\tau}\delta
    =-\fr{9c_{\rm D1}}{4\Mpl^2}\fr{2r-r_s}{r_s r}, \label{EOM_delta}
    \ee
which can be solved to yield
    \be
    \delta=-\fr{9c_{\rm D1}}{4\Mpl^2}\brb{2\fr{r}{r_s}-\ln\fr{r}{r_s}+C_0(\mR)}, \qquad
    \mR\coloneqq \bra{\fr{r}{r_s}}^{3/2}+\fr{3\tau}{2r_s}. \label{sol_delta}
    \ee
Here, $C_0$ is an arbitrary function corresponding to the homogeneous solution of \eqref{EOM_delta}.
In order for $\delta$ to be finite at the spatial infinity, the large-$\mR$ behavior of $C_0$ has to be of the form
    \be
    C_0(\mR)=-2\mR^{2/3}+\fr{2}{3}\ln \mR+\mO(\mR^0),
    \ee
which yields
    \be
    \delta=\fr{9c_{\rm D1}}{2\Mpl^2}\brb{\fr{\tau}{\sqrt{r_s r}}+\mO\bra{(r/r_s)^0}}.
    \label{sol_delta_inf}
    \ee
This result coincides with the one in \cite{Mukohyama:2005rw} for the case of ghost condensation, though we are now studying a different model of U-DHOST.
Note that, as can be seen in \eqref{sol_delta}, the variable~$\mR$ contains the spatial dependence, which was necessary to remove the divergent behavior of $\delta$ at the spatial infinity, as well as the time dependence, which yields the time dependence of $\delta$ in \eqref{sol_delta_inf}. 
This shows that the approximate solution must depend on time.
As we shall see in \S\ref{sec:BH_accretion}, the time-dependent solution for $\delta$ can be interpreted as the black hole mass growth due to the scalar field accretion, which puts a constraint on the value of $M$.

Let us consider the remaining EOMs, which can be reorganized as
    \begin{align}
    \bra{\pa_r-\sqrt{\fr{r}{r_s}}\pa_\tau}H_2
    &=-\fr{9c_{\rm D1}}{2\Mpl^2r}+\fr{9c_{\rm D1}}{4q^4F_{0XX}\,r^3}\bra{2+\fr{3q^2c_{{\rm D1},X}}{c_{\rm D1}}\fr{r_s}{r}}+\fr{\Mpl^2\sqrt{r_s}}{2q^4F_{0XX}}\pa_r\bra{r^{-3/2}\pa_\tau\delta}, \label{EOM_H2} \\
    H_1&=\fr{9c_{\rm D1}}{8q^4F_{0XX}\,r^2}\bra{1+\fr{q^2c_{{\rm D1},X}}{c_{\rm D1}}\fr{r_s}{r}}+\fr{\delta}{2}-\fr{\Mpl^2}{4q^4F_{0XX}}\sqrt{\fr{r_s}{r^3}}\,\pa_\tau\delta+\bra{1-\fr{r}{2r_s}}H_2. \label{EOM_H1}
    \end{align}
With the knowledge of the solution~\eqref{sol_delta} for $\delta$, one can first integrate \eqref{EOM_H2} to fix $H_2$, and then \eqref{EOM_H1} fixes $H_1$.
After the integration of \eqref{EOM_H2}, we obtain
    \be
    H_2=-\fr{9c_{\rm D1}}{4q^4F_{0XX}\,r^2}\brb{1+\fr{q^2c_{{\rm D1},X}}{c_{\rm D1}}\fr{r_s}{r}+\fr{3}{4}\sqrt{\fr{r}{r_s}}C_0'(\mR)+\fr{9r^2}{8r_s^2}\bra{\ln\fr{r}{r_s}} C_0''(\mR)}
    -\fr{9c_{\rm D1}}{2\Mpl^2}\brb{\ln\fr{r}{r_s}+C_1(\mR)},
    \ee
with $C_1$ being an arbitrary function.
By choosing $C_1(\mR)$ as
    \be
    C_1(\mR)=-\fr{2}{3}\ln\mR+\mO(\mR^{-2/3}),
    \ee
the perturbation variables~$H_1$ and $H_2$ remain finite at the spatial infinity:
    \be
    H_1=\fr{27c_{\rm D1}}{8\Mpl^2}\brb{\sqrt{\fr{r_s}{r^3}}\,\tau+\mO\bra{(r/r_s)^0}}, \qquad
    H_2=\fr{9c_{\rm D1}}{2\Mpl^2}\brb{\sqrt{\fr{r_s}{r^3}}\,\tau+\mO(r_s/r)}.
    \label{sol_H1H2_inf}
    \ee

For more generic higher-order scalar-tensor theories described by \eqref{HOST}, the system of EOMs yields many additional terms, including higher-derivative ones, compared to the simple case~\eqref{simple_example} of U-DHOST, and hence the above analysis does not directly apply.
Nevertheless, as we see below, the effect of the additional terms is suppressed and the solution is essentially described by the one for the simple model discussed above.
We assume that the coefficient functions in \eqref{HOST} and their derivatives satisfy the following scaling:
    \begin{equation}
    \begin{split}
    &X=\hat{X}M^4, \qquad
    F_{0XX}=\frac{\hat{F}_{0\hat{X}\hat{X}}}{M^4}, \qquad
    F_1=\hat{F}_1M, \qquad
    F_2=\fr{\Mpl^2}{2}+\hat{f}M^2, \qquad 
    F_{2X}=\frac{\hat{f}_{2\hat{X}}}{M^2}, \\
    &A_1=\frac{\hat{A}_1}{M^2}, \qquad
    A_2=\frac{\hat{A}_2}{M^2}, \qquad
    A_3=\frac{\hat{A}_3}{M^6}, \qquad
    A_4=\frac{\hat{A}_4}{M^6}, \qquad
    A_5=\frac{\hat{A}_5}{M^{10}}, \qquad
    A_{1X}=\frac{\hat{A}_{1\hat{X}}}{M^6},\qquad
    \cdots,
    \end{split} \label{scaling}
    \end{equation}
where all the hatted quantities are dimensionless and assumed to be of order unity.
Here, $M\,(\ll M_{\rm Pl})$ is the mass scale of derivative expansion which we regard as the EFT cutoff.
Then, one can verify that, up to the leading order in $M/\Mpl$ and $\epsilon=1/(M^2r_s^2)$, the configurations~\eqref{sol_delta_inf} and \eqref{sol_H1H2_inf} still solve the system of EOMs for generic higher-order scalar-tensor theories.
In this sense, the solutions~\eqref{sol_delta_inf} and \eqref{sol_H1H2_inf} would be universal in generic higher-order scalar-tensor theories so long as the scaling~\eqref{scaling} holds and $\epsilon=1/(M^2r_s^2)\ll 1$.

\section{Black hole accretion}\label{sec:BH_accretion}

Based on the time-dependent solution constructed in \S\ref{sec:approx_stealth}, we estimate the mass growth rate of the black hole due to the accretion of the scalar field, following the discussions in \cite{Cheng:2006us} in the case of (ungauged and gauged) ghost condensation. 
For this purpose, let us rewrite the solution~\eqref{sol_delta} for $\delta$, which amounts to the perturbation of the Misner-Sharp mass, in terms of the ingoing Eddington-Finkelstein coordinate~$v\coloneqq t+r_*$, with $r_*\coloneqq r+r_s\ln(r/r_s-1)$ being the tortoise coordinate.
We note that $\tau\simeq v$ and hence $\mR\simeq 3v/(2r_s)$ near the black hole horizon and in the far future when $v\gg r$ [see Eqs.~\eqref{t-to-tau} and \eqref{sol_delta}].
Therefore, for large $v$, the solution~\eqref{sol_delta} has the form
    \be
    \delta\simeq \fr{9\hat{c}_{\rm D1}M^2}{2\Mpl^2}\bra{\fr{3v}{2r_s}}^{2/3}, \label{delta_large-v}
    \ee
with $\hat{c}_{\rm D1}\coloneqq \cD/M^2$.
Provided that $\hat{c}_{\rm D1}>0$, this equation describes how the black hole mass increases at the late time.\footnote{As already stated before, in \cite{DeFelice:2022xvq}, it was shown that the condition~$\hat{c}_{\rm D1}>0$ is necessary for the scordatura mechanism to be successful.}

One can renormalize $r_s$ at each instance to discuss the time evolution of the black hole mass, which can be defined, in standard units, as
    \be
    2GM_{\rm BH}\coloneqq r_s(1+\delta),
    \ee
with $G=(8\pi \Mpl^2)^{-1}$ being the gravitational constant.
This implies
\begin{equation}
M_{{\rm BH}}^{2/3}=\left(\frac{r_{s}}{2G}\right)^{2/3}(1+\delta)^{2/3}
\simeq\left(\frac{r_{s}}{2G}\right)^{2/3}\left(1+\frac{2}{3}\delta\right),
\end{equation}
where one can substitute \eqref{delta_large-v} to yield
\begin{equation}
\left(\frac{2G}{r_{s}}\right)^{2/3}M_{{\rm BH}}^{2/3}
\simeq 1+\frac{3\hat{c}_{\rm D1}M^{2}}{\Mpl^{2}}\left(\frac{3}{2r_{s}}\right)^{2/3}v^{2/3}.
\end{equation}
Differentiating both sides of this equation with respect to $v^{2/3}$, we have
\begin{equation}
\left(\frac{2G}{r_{s}}\right)^{2/3}\frac{\D(M_{{\rm BH}}^{2/3})}{\D(v^{2/3})}
\simeq\frac{3\hat{c}_{\rm D1}M^{2}}{\Mpl^{2}}\left(\frac{3}{2r_{s}}\right)^{2/3},
\end{equation}
or equivalently,
\begin{equation}
\frac{\D(M_{{\rm BH}}^{2/3})}{\D(v^{2/3})} 
\simeq\frac{3\hat{c}_{\rm D1}M^{2}}{\Mpl^{2}}\left(6\pi\Mpl^{2}\right)^{2/3}.
\end{equation}
Then, we have
\begin{equation}
M_{{\rm BH}}^{2/3}
\simeq M_{{\rm BH},0}^{2/3}+ 
\frac{3\hat{c}_{\rm D1}M^{2}}{\Mpl^{2}}\left(6\pi\Mpl^{2}v\right)^{2/3},
\end{equation}
with $M_{{\rm BH},0}$ being the initial mass of the black hole.
This equation sets a constraint on $\hat{c}_{\rm D1}M^2$.

An example of stellar-mass black holes whose age can be estimated is in the X-ray binary~XTE J1118+480.
The black hole has a mass of $M_{\rm BH}\sim 7.24M_{\odot}$~\cite{Cherepashchuk:2019qap} and its age is estimated to be either $\sim240~{\rm Myr}$ or $\sim7~{\rm Gyr}$ depending on possible formation processes of the binary~\cite{Mirabel:2001ay}.
Note that we have $M_{{\rm BH},0}\gtrsim3M_{\odot}$ from the theory of stellar evolution.
If the black hole is described by our solution, the mass growth due to the accretion of the scalar field must not exceed the difference between the observed mass~$M_{\rm BH}$ and the initial mass~$M_{{\rm BH},0}$.
In order to be conservative, we employ the younger age of $v\simeq t\sim 240~{\rm Myr}$ to obtain
    \be
    \fr{3\hat{c}_{\rm D1}M^2}{\Mpl^2}\bra{6\pi\Mpl^2\times240~{\rm Myr}}^{2/3}
    \lesssim \bra{7.24M_{\odot}}^{2/3}-\bra{3M_{\odot}}^{2/3}.
    \ee
This provides an upper bound on $M$ as
    \be
    \hat{c}_{\rm D1}^{1/2}M
    \lesssim 2\times10^{11}~{\rm GeV}. \label{eq:cutoff}
    \ee
As we mentioned at the end of \S\ref{sec:approx_stealth}, for this approximate solution to be trusted, one requires that $\epsilon=1/(M^2r_s^2)\ll 1$, or $M\gg \Mpl^2/M_{\rm BH}$.
For this particular astrophysical black hole, this corresponds to setting 
$M\gg 10^{-21}$~GeV, which is well satisfied for a high-energy EFT cutoff scale consistent with the bound in \eqref{eq:cutoff}.

\section{Conclusions}\label{sec:conc}

Scalar-tensor theories have been studied quite in detail as they represent simple theoretical extensions from the concordance $\Lambda$CDM model. From the seminal work of Horndeski, the field of scalar-tensor theories has been continuously expanding, trying to find new sensible extensions which might have the non-trivial phenomenology required to describe and solve today's puzzles/tensions in cosmology. In particular, the Horndeski condition on the Euler-Lagrange equations to be of the second order has been proved to be non-necessary to have a healthy theory without the Ostrogradsky ghost. This led to even more general kinds of ghost-free scalar-tensor theories. These theories typically need to satisfy degeneracy conditions for the higher-derivative terms in order to remove the Ostrogradsky mode, and such theories were named DHOST. Other theories, forming the U-DHOST framework, satisfy degeneracy conditions only when they are evaluated on the unitary gauge where the scalar field is spatially uniform. These theories possess a shadowy mode which satisfies a three-dimensional elliptic differential equation for which appropriate boundary conditions need to be imposed. 
Having said that, the degeneracy conditions are not protected in general by symmetry, and hence they are expected to be detuned by quantum corrections.
The detuning of the degeneracy conditions revives the Ostrogradsky ghost, but it is harmless if its mass is above the cutoff scale of the theory, which we denote by $M$.
Therefore, higher-order scalar-tensor theories away from the (U-)DHOST class should not be thought of as excluded a priori, and they should instead be considered as EFTs which can be applied, for instance, to cosmology.

In this paper, we investigated (approximately) stealth solutions in the healthy higher-order scalar-tensor theories mentioned above. By stealth solutions, we mean solutions where the metric configuration exactly matches the GR counterpart, whereas the scalar field shows a non-trivial profile. These solutions evidently are interesting as they may evade solar system constraints, if the metric is assumed to be the Schwarzschild(-de Sitter) one. We found that, unless the theory is of DHOST type, the Schwarzschild(-de Sitter) configuration cannot be an exact solution for the EOMs. On the other hand, in the context of DHOST theories, it is known that perturbations around exactly stealth solutions are infinitely strongly coupled~\cite{Babichev:2018uiw,deRham:2019gha,Motohashi:2019ymr,Khoury:2020aya,Takahashi:2021bml}. Therefore, if we are to consider stealth or almost stealth solutions, then we need to consider theories beyond DHOST including scordatura terms, such as ghost condensation and U-DHOST, and as a result the solutions are not exactly stealth but only approximately stealth. Indeed, we found that, if we consider $M\gg\Mpl^2/M_{\rm BH}$ (a condition which holds true for a typical black hole mass~$M_{\rm BH}$ and a high-energy cutoff scale below Planck, i.e., $M\ll\Mpl$), then the solution for the metric differs from the Schwarzschild one by terms that are suppressed by the factor~$M^2/\Mpl^2$. Furthermore, as in the case of ghost condensation~\cite{Mukohyama:2005rw}, 
the black hole mass slowly grows in time, which can be understood as the accretion of the scalar field.
This growth can be constrained if we know the initial and final masses and the age of a black hole. On considering this growth due to the scalar field accretion, we find a constraint on the cutoff scale of the theory~$\hat{c}_{\rm D1}^{1/2} M\lesssim 2\times 10^{11}$~GeV, 
where $\hat{c}_{\rm D1}$ is, in general, a dimensionless parameter of order unity.

We believe our approximately stealth solution can provide the necessary playground to study these theories and their phenomenology on astrophysical scales. It would be interesting to further study the stability of such solutions, where a careful treatment of the Ostrogradsky/shadowy mode would be required.
We defer this analysis to a future project of ours.


\acknowledgments{
The authors are thankful to Felix Mirabel for pointing out to us updated values for the mass of the X-ray binary XTE J1118+480. 
The work of A.D.F.\ was supported by JSPS (Japan Society for the Promotion of Science) Grants-in-Aid for Scientific
Research No.\ 20K03969.
K.T.\ was supported by JSPS KAKENHI Grant No.\ JP21J00695. 
The work of S.M.\ was supported in part by JSPS Grants-in-Aid for Scientific Research No.\ 17H02890, No.\ 17H06359, and by World Premier International Research Center Initiative, MEXT, Japan. 
}


\bibliographystyle{mybibstyle}
\bibliography{bib}

\end{document}